\documentclass[%
reprint,
superscriptaddress,
 amsmath,amssymb,
 aps,
]{revtex4-2}

\usepackage{graphicx}% Include figure files
\usepackage{dcolumn}% Align table columns on decimal point
\usepackage{bm}% bold math
\usepackage{siunitx}  				% SI-Einheiten
\usepackage{blindtext}
\usepackage{xcolor}% for defining own colors
\usepackage{jabbrv}
\definecolor{cblau}{rgb}{0.69,0.89,1.0}			% blau
\definecolor{tumblauTitel}{cmyk}{1.0,0.43,.0,.1} % TUM-blau
\definecolor{ao}{rgb}{0.0, 0.0, 1.0} %pure blue
\definecolor{Gray}{gray}{0.9}
\newcommand{\moly}{$\text{MoS}_2$}

\newcommand{\wese}{$\text{WSe}_2$}

\newcommand{\molyd}{$\text{MoS}_2~$}

\newcommand{\sildiox}{$\text{SiO}_2$}

\newcommand{\Bperp}{$B_{\perp}$}
\newcommand{\Bperpd}{$B_{\perp}~$}
\newcommand{\gperp}{$\text{g}_{\perp}$}

\newcommand{\Bpar}{$B_{\parallel}$}
\newcommand{\Bpard}{$B_{\parallel}~$}

\newcommand{\circmin}{$\sigma^{-}$}
\newcommand{\circmind}{$\sigma^{-}~$}
\newcommand{\circplus}{$\sigma^{+}$}
\newcommand{\circplusd}{$\sigma^{+}~$}

\newcommand{\muw}[1]{\SI{#1}{\micro\watt}}
\newcommand{\K}[1]{\SI{#1}{\kelvin}}
\newcommand{\nm}[1]{\SI{#1}{\nano\meter}}
\newcommand{\V}[1]{\SI{#1}{\volt}}
\newcommand{\kV}[1]{\SI{#1}{\kilo\volt}}

\newcommand{\eV}[1]{\SI{#1}{\electronvolt}}
\newcommand{\mum}[1]{\SI{#1}{\micro\meter}}
\newcommand{\meV}[1]{\SI{#1}{\milli\electronvolt}}

\newcommand{\perc}[1]{\SI{#1}{\percent}}

\newcommand{\angstrom}[1]{\SI{#1}{\angstrom}}

\newcommand{\nW}[1]{\SI{#1}{\nano\watt}}

\newcommand{\mueVTsq}[1]{\SI{#1}{\micro\electronvolt\text{ }\tesla^{-2}}}
\newcommand{\T}[1]{\SI{#1}{\tesla}}

\usepackage{ragged2e}
\usepackage{caption}
\usepackage[caption2]{ccaption} % Continued captions package
\DeclareCaptionLabelSeparator{dash}{ $\mathbf{|}$ }
\DeclareCaptionJustification{blocktext}{\justifying}

\begin{document}

\preprint{APS/123-QED}

\title{Spin-defect characteristics of single sulfur vacancies in monolayer \moly}

\author{A. Hötger}
\affiliation{Walter Schottky Institute and Physics Department, TU Munich, 85748 Garching, Germany.}

\author{T. Amit}
\affiliation{Department of Molecular Chemistry and Materials Science, Weizmann Institute of Science, Rehovot, Israel.}

\author{J. Klein}
\affiliation{Department of Materials Science and Engineering, Massachusetts Institute of Technology, Cambridge, Massachusetts 02139, USA}
%\affiliation{Walter Schottky Institut and Physics Department, TU Munich, 85748 Garching, Germany.}

\author{K. Barthelmi}
\affiliation{Walter Schottky Institute and Physics Department, TU Munich, 85748 Garching, Germany.}

\author{T. Pelini}
\affiliation{Laboratoire National des Champs Magnetiques Intenses, CNRS-UGA-UPS-INSA-EMFL, 38042 Grenoble, France.}

\author{A. Delhomme}
\affiliation{Laboratoire National des Champs Magnetiques Intenses, CNRS-UGA-UPS-INSA-EMFL, 38042 Grenoble, France.}

\author{S. Rey}
\affiliation{Department of Photonics Engineering, Technical University of Denmark, 2800 Kgs. Lyngby, Denmark.}

\author{M. Potemski}
\affiliation{Laboratoire National des Champs Magnetiques Intenses, CNRS-UGA-UPS-INSA-EMFL, 38042 Grenoble, France.}
\affiliation{Institute of Experimental Physics, Faculty of Physics, University of Warsaw, 02-093 Warszawa, Poland.}

\author{C. Faugeras}
\affiliation{Laboratoire National des Champs Magnetiques Intenses, CNRS-UGA-UPS-INSA-EMFL, 38042 Grenoble, France.}

\author{G. Cohen}
\affiliation{Department of Molecular Chemistry and Materials Science, Weizmann Institute of Science, Rehovot, Israel.}

\author{D. Hernangómez-Pérez}
\affiliation{Department of Molecular Chemistry and Materials Science, Weizmann Institute of Science, Rehovot, Israel.}

\author{T. Taniguchi}
\affiliation{International Center for Materials Nanoarchitectonics, National Institute for Materials Science,  1-1 Namiki, Tsukuba 305-0044, Japan.}

\author{K. Watanabe}
\affiliation{Research Center for Functional Materials, National Institute for Materials Science, 1-1 Namiki, Tsukuba 305-0044, Japan.}

\author{C. Kastl}
\affiliation{Walter Schottky Institute and Physics Department, TU Munich, 85748 Garching, Germany.}

\author{J.J. Finley}
\affiliation{Walter Schottky Institute and Physics Department, TU Munich, 85748 Garching, Germany.}

\author{S. Refaely-Abramson}
\affiliation{Department of Molecular Chemistry and Materials Science, Weizmann Institute of Science, Rehovot, Israel.}

\author{A.W. Holleitner}
\affiliation{Walter Schottky Institute and Physics Department, TU Munich, 85748 Garching, Germany.}

\author{A.V. Stier}
\affiliation{Walter Schottky Institute and Physics Department, TU Munich, 85748 Garching, Germany.}

\date{\today}% It is always \today, today,

\begin{abstract}
    Single spin-defects in 2D transition-metal dichalcogenides are natural spin-photon interfaces for quantum applications. Here we report high-field magneto-photoluminescence spectroscopy from three emission lines (Q1, Q2 and Q*) of He-ion induced sulfur vacancies in monolayer \moly. Analysis of the asymmetric PL lineshapes in combination with the diamagnetic shift of Q1 and Q2 yields a consistent picture of localized emitters with a wave function extent of $\sim$\nm{3.5}. The distinct valley-Zeeman splitting in out-of-plane $B$-fields and the brightening of dark states through in-plane $B$-fields necessitates spin-valley selectivity 
    of the defect states and lifted spin-degeneracy at zero field. Comparing our results to ab-initio calculations identifies the nature of Q1 and Q2 and suggests that Q* is the emission from a chemically functionalized defect. Analysis of the optical degree of circular polarization reveals that the Fermi level is a parameter that enables the tunability of the emitter. These results show that defects in 2D semiconductors may be utilized for quantum technologies.
    
\end{abstract}

\maketitle

%\tableofcontents

\section{\label{Introduction}Introduction}

    Spin-defects in host crystals can be fundamental building blocks for quantum technologies, such as computing, sensing or communication\cite{Reserbat-Plantey2021,Gottscholl2021,Turunen2022, Tarasenko2018}.
    %They introduce a variety of unique properties on the atomic-scale and represent the ultimate limit of nanotechnologies. 
    For instance, color centers in diamond have been investigated since the early 1980s, of which the nitrogen vacancy (NV) center is the most prominent example\cite{Gruber1997,Doherty2013}. In this defect, the crystal field splitting lifts the ground state spin degeneracy and provides the required unique quantum degree of freedom to form an addressable two-level system\cite{Dobrovitski2013,Casola2018,Bradac2019,Hernandez-Gomez2021}. % Extremely long spin coherence times, addressability and immensity of coherent control make the negatively charged nitrogen-vacancy (NV) center the most promising type of defect in diamond for solid-state spin qubits. \cite{Dobrovitski2013, Doherty2013, Hernandez-Gomez2021} 
    %Additional to numerous applications as quantum sensors on the atomic-scale \cite{Dobrovitski2013, Doherty2013, Hernandez-Gomez2021}, 
    Additionally, NV centers are single photon sources\cite{Draebenstedt1999, Brouri2000, Kurtsiefer2000} and therefore constitute excellent building blocks for future quantum photonic circuits.
    However, a key prerequisite for such applications is the ability to position defects deterministically. This is a challenge for defects in 3D crystals, such as single NV centers, as they can be positioned either vertically or laterally with high precision, but not both simulateneously\cite{Ohno2012,Martin1999,Lesik2013,Pezzagna2010}. This disadvantage can be overcome by creating optically addressable spin-defects in 2D host crystals. Localized single photon emission in 2D materials was first discovered in monolayer \wese, a prototypical member of the semiconducting 2D transition metal dichalcogenides (TMDs)\cite{He2015,Chakraborty2015,Srivastava2015,Koperski2015, Tonndorf2015}. Subsequently, single photon emitters were discovered in hexagonal boron nitride (hBN)\cite{Tran2016}.
    Contrasted with hBN, TMDs have strong light-matter coupling\cite{Mak2010} and locked spin-valley physics\cite{Xiao2012}, which provides a natural spin-photon interface. Moreover, the 2D semiconducting host crystal has enabled new possibilities to engineer and manipulate these defects\cite{Lin2016, Chakraborty2015, Mukherjee2020, Hotger2021}, which led to further advances in quantum devices, such as quantum light emitting diodes\cite{Palacios-Berraquero2016, Schwarz2016, Clark2016}.
    
    First approaches for the deterministic creation of quantum emitters in 2D materials made use of strain potentials, for instance induced by a textured substrate\cite{Kumar2015, Kern2016, Branny2017, Palacios-Berraquero2017, Branny2016, Proscia2018,Parto2021}. This results in a local bandstructure modulation in the host crystal, limited by the bending radius of the material, yet the latter approach intrinsically lacks reproducibility. Furthermore, the confining potential often breaks crystal symmetries, leading to the loss of valley optical selection rules. A higher degree of spatial resolution and reproducibility can be achieved by using the accuracy of electron-beam or focused ion beam irradiation\cite{Komsa2012, Moody2018, Klein2019, Fournier2021, Kretschmer2018}. Specifically, He-ions can be precisely focused and create optically active point defects in monolayer \molyd\cite{Klein2019} with a precision better than \nm{10}\cite{Mitterreiter2020}. In photoluminescence (PL) spectroscopy, spectrally narrow emission lines appear about \meV{200} red-shifted from the neutral exciton of He-ion irradiated \molyd\cite{Klein2019}. Second order correlation measurements unambiguously showed single photon emission from single He-ion irradiation sites, which in turn could be related to the generated point defects\cite{Barthelmi2020, Klein2021}.
    A specific advantage is that these defects can be implanted into more complex, electronic device heterostructures, allowing for the electrical control of quantum emission.\cite{Hotger2021}.
    
    The microscopic origin of various localized emission centers is currently a matter of debate. %For example, the presence of atomic defects in TMDs leads to modifications in the electronic band structure. \cite{Li2019a} 
    One specific defect complex, which is predominant in He-ion irradiated \moly, is the chalcogen vacancy, where one sulfur atom has been removed from the host lattice\cite{Mitterreiter2020}. Sub-gap quantum emission from this defect\cite{Refaely-Abramson2018} was suggested to originate from a relaxation cascade where an optical interband excitation creates a bound electron-hole pair that subsequently localizes into the defect and radiatively recombines\cite{Mitterreiter2021}. However, pristine defect states have been predicted to be essentially spin degenerate\cite{Refaely-Abramson2018, Gupta2019}. Moreover, other relaxation pathways, such as defect-to-band transitions are in principle possible, and, due to the strong spin-orbit interaction in the host \moly, considerations with respect to spin-valley physics have yet to be taken into account.
    In this manuscript, we investigate three distinct emission lines of He-ion induced sulfur vacancies created in monolayer \molyd by high-field magneto-optical spectroscopy, which has previously been shown to be an important tool to investigate the excitonic spin-valley physics in 2D TMDs\cite{Stier2016,Goryca2019,Li2020}.
    We identify the bands involved in the optical quantum emission and show that an energy dependent degree of hybridization between atom-like defect states and the \molyd bandstructure leads to varying degree of valley selectivity for the distinct electron-hole transitions. Our results display that sulfur vacancies in monolayer \molyd are spin-defects that can be tailored to specific quantum applications.    

\section{\label{Results}Results}
	\subsection{\label{Quantum_emission_in_mos2} Photoluminescence from sulfur vacancies in monolayer \moly}

	    \begin{figure*}
			\includegraphics{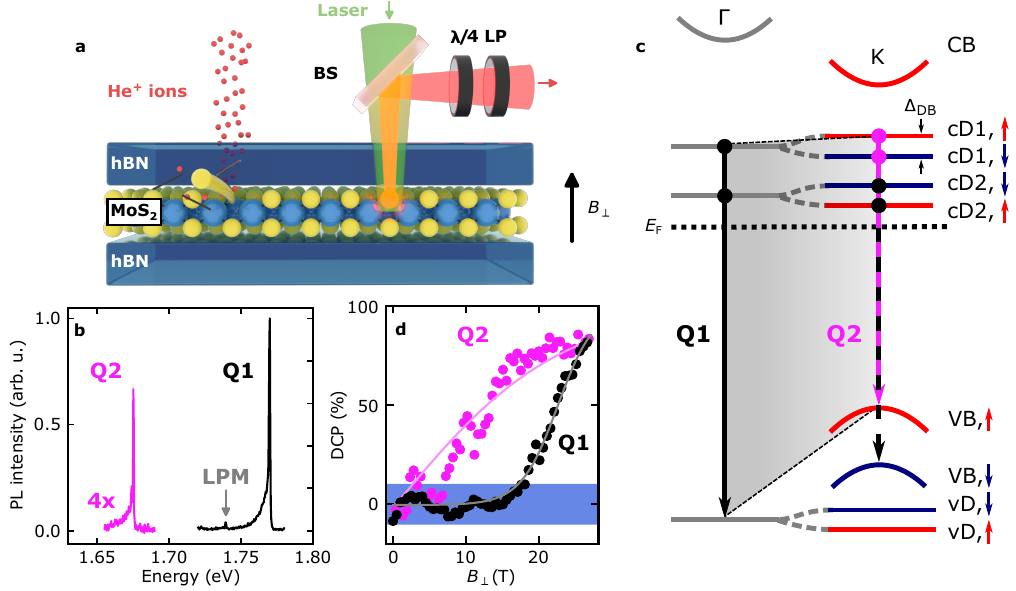}
% 			\internallinenumbers
			\caption{\label{Quantum_emission_overview}\textbf{Defect luminescence of He-ion irradiated monolayer \moly.} a) Sketch of the monolayer \molyd encapsulated in hexagonal boron nitride (hBN) illustrating the He-ion irradiation and the out-of-plane magneto-spectroscopy scheme. b) Typical low-temperature photoluminescence (PL) spectra of the quantum emission Q1, Q2, and the local phonon mode (LPM) of Q1. c) Illustration of the \molyd bandstructure at the $K$ and $\Gamma$ valleys including the defect states of a sulfur vacancy. The modified bandstructure shows flat defect-related levels vD, cD2 and cD1. The red (blue) color represents the spin-up (spin-down) eigenstate of the associated band. A small spin splitting of the defect states is observed in our calculations at the $K$/$K'$ valleys, as discussed in the main text. At \Bperp $= \T{0}$, the Fermi energy $E_\text{F}$ lies below the unoccupied defect states. Solid circles mark significant defect states. d) Degree of circular polarization (DCP) versus \Bperpd for Q1 and Q2. The blue shaded area highlights the $\pm$\perc{10} experimental uncertainty. The oscillations in the DCP below $\sim$ \T{15} are predominantly due to fringe field induced Faraday rotation in the low temperature objective. The solid lines are fits to the data with equation (\ref{DCP_fit}) for Q1 and a Boltzmann statistics fit for Q2 (see Supplementary Figure 5).}
% 			The insets show isosurfaces of the quasiparticle wave function cD2 (in the bandgap) and vD (below the valence band) associated with the defect states in \moly. The wave functions of cD1 and cD2 are similar.}
		\end{figure*}
		
		The left hand side of Fig. \ref{Quantum_emission_overview}a shows the schematic of the sample under investigation. A monolayer \molyd is encapsulated in hBN, fabricated by standard dry viscoelastic stamping methods (see Methods for details). Subsequently, a focused He-ion beam is scanned across the sample, creating predominantly sulfur vacancies in the \moly\cite{Mitterreiter2020}. Our sample is He-ion irradiated with a pitch of $\sim$ \mum{2}, which allows us to selectively investigate individual irradiated locations (see Supplementary Figure 1a).
		Typical defect PL spectra at \K{1.7} and zero magnetic field are shown in Fig. \ref{Quantum_emission_overview}b. The dominant feature at \eV{1.75}, labeled Q1, has previously been identified as a single photon emitter\cite{Klein2021, Barthelmi2020} associated with an unpassivated sulfur vacancy\cite{Mitterreiter2021}. In this manuscript, we discuss only those locations which contained a single Q1 line, generally the case for the sample under investigation. The low energy tail of Q1 is attributed to the coupling of a localized state to acoustic phonons in \moly. Analysis of the lineshape with an independent boson model allowed the determination of the effective Bohr radius of this localized state to be $\approx$ 2-\nm{3}\cite{Klein2019}. A weak secondary feature about \meV{30} red-shifted from the zero phonon line (ZPL) of Q1, can be assigned to a phonon replica due to a local phonon mode (LPM) of this defect center\cite{Klein2021}. Emission line Q2 forms in a distinct energy band $\approx$ \meV{75} red-shifted from Q1, while another emission line Q* appears $\approx$ \meV{50} blue-shifted. Both lines generally appear fainter as compared to Q1, while the lineshape of all features are similar. The statistical evaluation of all emission lines investigated throughout the sample clearly indicates these three inhomogeneously broadened, yet distinct, emission bands (see Supplementary Figure 1c). Our zero B-field spectroscopy therefore establishes the observed emission lines to originate from localized defect centers. This is consistent with predictions of six spinor wave functions due to an atomistic defect associated with a sulfur vacancy resulting from its $C_{3v}$ symmetry, where two spin-split bands are above the Fermi energy and one is below it. (see Supplementary Figure 10).
% 		Figure \ref{Quantum_emission_overview}d shows calculated isosurfaces of the square of the wave functions associated with the electronic defect levels cD1, cD2 and vD, that are localized around the defect site. The wave functions are primarily composed of transition metal d-orbitals. \cite{Refaely-Abramson2018}
		As sketched in Fig. \ref{Quantum_emission_overview}c and discussed in detail below, we unambiguously identify Q1 as an excitonic transition predominantly between defect induced states (cD1/cD2 $\leftrightarrow$ vD) at the $\Gamma$ point, with significant hybridization across the Brillouin zone and specifically at the $K/K'$ points\cite{Refaely-Abramson2018}. This spread in momentum-space originates from the localized character of the sulfur vacancy, as shown by the calculated wave function distributions associated with the electronic defect levels cD1, cD2 and vD (see Supplementary Figure 10). The wave functions are primarily composed of transition metal d-orbitals and therefore contain the spin-valley physics of monolayer \molyd throughout the Brillouin zone\cite{Refaely-Abramson2018}.
	    For the energetically lower lying emission Q2, we identify the superposition of transitions between both spin-up/down defect induced conduction band states (cD1) and the \molyd valence band (VB) at the $K/K'$-points in the Brillouin zone. Although Q2 is also excitonic, the transitions are confined to the $K/K'$-points. We further demonstrate that Q* originates from a chemically functionalized sulfur vacancy and the emission is of character Q2.
		One of the central aspects of this manuscript is the observation of the valley dichroism of all emission lines through the valley Zeeman splitting and optical degree of circular polarization (DCP) in high magnetic fields. We note that valley selectivity stems from contributions at the $K/K'$-points to the excitonic transitions, which we probe via circular polarization resolved magneto-spectroscopy. As an example, the DCP of Q1 and Q2 are shown in Fig. \ref{Quantum_emission_overview}d, which, for Q1 reveals essentially no polarization in the B-field range below \T{15}, and a rapid rise of the DCP, tending towards unity at the highest fields. Q2 polarizes already at low fields. The observation that these emission lines show valley dichroism at finite magnetic field necessitates the lifting of the spin degeneracy, and we further proof in detail below that the spin degeneracy of the defect states in the $K/K'$ valleys is already lifted at zero magnetic field. Thus, the sulfur vacancy in \molyd can be considered as a spin-defect.
		        
		%Das Independent model nimmt die effective masse des elektrons m_e_eff = 0.44*m_e und eine effective masse des Lochs m_h_eff = 0.69*m_e an. (i.e. reduced mass of 0.27 m_e)
	\subsection{\label{Faraday_geometry}Out-of-plane magnetic field measurements on defects in \moly}
	  
		\begin{figure*}
			\includegraphics{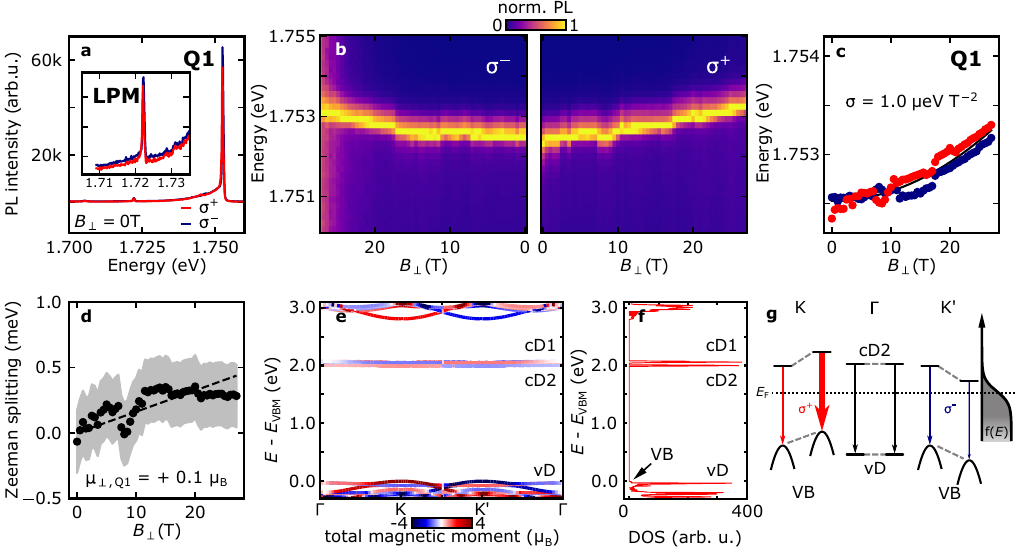}
% 			\internallinenumbers
			\caption{\label{Q1_LPM_Faraday}\textbf{Out-of-plane magnetic field \Bperpd dependent photoluminescence of defect luminescence Q1.} a) Low-temperature photoluminescence (PL) spectra of defect luminescence Q1 at zero out-of-plane magnetic field (\Bperp ) for \circplusd (red) and \circmind (blue) polarized detection. The zero-phonon line (ZPL) of Q1 occurs at \eV{1.752} with a red-shifted (\meV{30}) local phonon mode (LPM, see inset). b) PL versus \Bperpd for Q1. The left (right) panel shows the \circmind (\circplus ) polarized signal. The spectra were normalized to their maximum intensity. c) The fitted position of the ZPL of Q1 shows a diamagnetic shift of $1.0 \pm \mueVTsq{0.1}$. d) Valley Zeeman splitting of Q1 versus \Bperp. The black dashed line shows a fit to the data with a Zeeman splitting of $\mu_{\perp\text{,Q1}} = +0.1 \pm 0.1\ \mu_\text{B}$. e) Bandstructure of \molyd with a periodic sulfur-vacancy extracted from DFT calculations. The color code denotes the total magnetic moment at each k-point. f) Density of states (DOS) plot of the Brillouin zone, showing energetically narrow densities at the defect levels. g) Illustration of the two spin-split electron states in the $K/K'$ valley and possible optical transitions of Q1 for zero and finite magnetic field.}
		\end{figure*}
		
		 In order to investigate the nature of the observed emission bands, we employ magneto-PL spectroscopy, where the magnetic field \Bperpd up to \T{27} is applied perpendicular to the 2D sample plane and parallel to the optical beam path (Faraday geometry). The sample was mounted in a He-exchange gas cryostat with a bath temperature of $T_\text{Bath}=\K{4.2}$. The sample was excited with a linearly polarized continuous wave (CW) laser at a wavelength of \nm{515} and power of $\sim$ \muw{10}, focused to a beam spot of $\sim$ \mum{1}. The linear polarization excites interband transitions in both $K/K'$ valleys of the host \moly. At each positive magnetic field, we probe the circular dichroism by detecting the PL for \circmind and \circplusd polarization, which we calibrate with the well known valley Zeeman splitting of the neutral exciton in \molyd (see Supplementary Figure 2). As such, we minimize the impact of positional sample drift with respect to the beam path in very high magnetic fields. From the faint appearance of the negatively charged trion in the PL spectra and the value of the valley Zeeman splitting for the neutral exciton ($\mu_{\perp} = -2.8 \pm 0.1\ \mu_\text{B}$), we conclude that our \molyd crystal is weakly electron doped $n \approx \SI{5e11}{cm^{-2}}$\cite{Klein2021a}. Figure \ref{Q1_LPM_Faraday}a depicts typical polarization resolved PL spectra of Q1 and the LPM at \Bperpd $= \T{0}$. Unlike strain induced quantum emitters in monolayer TMDs\cite{Kumar2015, Branny2016, Yu2021}, the He-ion-induced defects show no valley dichroism at zero magnetic field (see Fig. \ref{Quantum_emission_overview}d)\cite{Mitterreiter2021}. This is expected for the $C_{3v}$ symmetry of an unperturbed sulfur vacancy with defect-to-defect transitions at the $\Gamma$-point.\cite{Refaely-Abramson2018,Gupta2019} The left (right) panel of Fig. \ref{Q1_LPM_Faraday}b shows the normalized PL of Q1 versus \Bperpd for \circmind (\circplus ) polarized detection. The position of the ZPL exhibits a monotonic blue-shift with increasing magnetic field for both polarizations. We plot the PL peak position in Figure \ref{Q1_LPM_Faraday}c and find that the average peak position for \circplusd and \circmind detection $\frac{1}{2}( E_{\sigma^{+}} + E_{\sigma^{-}} )$ can be fitted with a quadratic function ($\sim$ \Bperp$^2$) with a prefactor of $\sigma = 1.0\ \pm$ \mueVTsq{0.1}. The quadratic-in-\Bperpd blueshift is consistent with the expected diamagnetic shift of a bound particle in 2D, $\Delta E_\text{dia}=e^{2}\langle r^{2} \rangle B_\perp^{2}/8 m_\text{r}$\cite{Stier2016,Goryca2019}. The root mean square radius in the plane of the 2D material is expressed as $r_\text{rms} = \sqrt{8m_{r}\sigma}/e$, where $m_{r}$ is the reduced mass of the particle and $e$ the elementary charge, respectively. The observed diamagnetic shift coefficient of Q1 is roughly $5\times$ larger than that of the neutral \molyd exciton\cite{Stier2016,Goryca2019}. Assuming the reduced mass for the neutral exciton and Q1 are the same, ($m_\text{r} = 0.275\ m_\text{0}$\cite{Goryca2019}), the observed diamagnetic shift yields a particle size $r_\text{rms}=\nm{3.5}$, consistent with the findings of the independent boson model discussed above. 
		 The valley Zeeman splitting, defined from the difference $E_{\sigma^{+}} - E_{\sigma^{-}} = \mu_\perp B$ is shown in Fig. \ref{Q1_LPM_Faraday}d. In contrast to other quantum emitters in 2D materials\cite{Srivastava2015, He2015, Koperski2015, Chakraborty2015,Branny2016, Brotons2019, Lu2019}, the Q1 emission shows little, but experimentally detectable, positive valley Zeeman splitting of $\mu_{\perp\text{,Q1}} = +0.1 \pm 0.1\ \mu_\text{B}$. Such a vanishing Zeeman splitting has recently been observed on a quantum emission and was attributed to a quasiparticle transition between pristine conduction band and in-gap defect state\cite{Dang2020}. However, the latter study does not provide a full evaluation on the excitonic effects, which are crucial for magneto-spectroscopy in TMDs.
		 In order to get a first insight into the nature of the Q1 emission line, Figure \ref{Q1_LPM_Faraday}e shows the DFT bandstructure of \molyd with a $\perc{2}$ sulfur vacancy density. 
		 The pristine bandstructure of \molyd is essentially unaffected by the presence of the sulfur vacancies. However, additional electronic states lie within the bandgap (cD1, cD2), as well as in the valence band (vD) of \moly. These states are relatively flat in k-space, which yields a high joint density of states for defect-to-defect transitions in this system (see Fig. \ref{Q1_LPM_Faraday}f), particularly at the $\Gamma$-point. For a pure defect-to-defect transition at the $\Gamma$-point, we expect exactly zero valley Zeeman splitting due to the vanishing of the valley selectivity. In turn, the finite valley Zeeman splitting is consistent with a Q1 emission dominated by defect-to-defect transitions at the $\Gamma$-point and its hybridization with defect-to-band transitions at the $K/K'$-points, which we discuss in detail below. 
		Although we find $\mu_{\perp\text{,Q1}} \approx 0$, we measure a large degree of circular polarization (see Fig. \ref{Quantum_emission_overview}d) at high magnetic fields, calculated from the integrated PL intensities of both helicities $(I_{\sigma^{+}} - I_{\sigma^{-}})/(I_{\sigma^{+}} + I_{\sigma^{-}})$. 
		This $B$-field induced circular polarization requires spin polarized states to participate in the optical transition. 
		The measured DCP at \Bperpd = \T{0} is within the experimental uncertainty of $\pm$\perc{10} (indicated by the shaded area in Fig. \ref{Quantum_emission_overview}d). The uncertainty originates from spectral jitter as well as uncompensated Faraday rotation of the linearly polarized excitation light combined with imperfectly aligned $\lambda/4$ and linear polarizers in the detection path. In our understanding, the intensities of the \circplus- and  \circmin-polarized transition in either the $K$ or the $K'$-valley are weighted with the probability of the defect level at $E_\text{cD2}$ to be occupied by an electron of the Fermi sea using the Fermi-Dirac $f_\text{FD}$ distribution. In turn, we assume following expression to fit the DCP as a function of the applied magnetic field.
		
        \begin{multline}
            \text{DCP}(B) = \\
           =\dfrac{f_\text{FD}(E_\text{cD2,0} + \mu_\text{cD2} \cdot B) - f_\text{FD}(E_\text{cD2,0} - \mu_\text{cD2} \cdot B)}{f_\text{FD}(E_\text{cD2,0} + \mu_\text{cD2} \cdot B) + f_\text{FD}(E_\text{cD2,0} - \mu_\text{cD2} \cdot B)},
            \label{DCP_fit}
        \end{multline}

         with the Zeeman shift of the cD2 states in the $K/K'$ valleys to be $\mu_\text{cD2}\cdot B$ and $E_\text{cD2,0}$ the energy of state cD2 at zero field. Fitting the data of Q1 in Fig. \ref{Quantum_emission_overview}d with equation (\ref{DCP_fit}) (see line in Fig. \ref{Quantum_emission_overview}d) yields $\mu_\text{cD2} = 2.6 \pm 0.5 \mu_\text{B}$, with the error given by the systematic error of our polarization alignment in the high-magnetic field setup. Moreover, the fit assigns cD2 to be above the Fermi energy $E_\text{F}$ at \Bperpd $=$ \T{0}, with $E_\text{cD2} - E_\text{F} = 3.2 \pm$ \meV{0.7}. We note that the ab-initio calculations as in Fig. \ref{Q1_LPM_Faraday}e determine $\mu_\text{cD2}$ to be $0.86 \mu_\text{B}$, which is again clearly positive. For the experimental (ab-initio) value, the Zeeman shift of cD2 at \T{22.5} equals $\sim$\meV{3.4} (\meV{1.12}), which is larger than the thermal energy of $\sim$\meV{0.7} at \K{4.2}. In our understanding, this explains that the DCP can be detected at high magnetic fields at the given temperature. We note that this interpretation is corroborated by a gate-tunable device where the DCP changes sign by reversing the polarity of \Bperpd (see Supplementary Figure 4). As a consequence, an applied magnetic field lifts the spin degeneracy of Q1.
         Moreover, a characteristic blue-shift of the quantum emission with increasing charge carrier density suggests the defects to be charge neutral (see Supplementary Figure 4).
         Based on these findings, Figure \ref{Q1_LPM_Faraday}g summarizes the interpreted bandstructure at $K/K'$ and $\Gamma$ for zero and finite \Bperp. With increasing magnetic fields, the spin-up state of cD2 in the $K$ valley (lowest defect state in the bandgap at the $K$-point) is pushed away from the Fermi edge, which decreases the possibility for it to be occupied with an electron from the Fermi edge. This increases the part of \circplusd polarized light emitted at the $K$-point and eventually polarizes the overall emission. Conversely, at the $K'$-point, the spin-down defect state is the lowest defect state in the bandgap and is pushed towards the Fermi edge. Therefore, the intensity of \circmind polarized light subsequently diminishes. 
         From the combined experimental observations of small valley Zeeman splitting and strongly \Bperp-dependent DCP, we identify Q1 as an excitonic emission of a neutral sulfur vacancy in \molyd between hybridized defect-to-defect and defect-to-band transitions and relate the observed DCP to a combination of magnetic field induced Zeeman shifts and occupation effects. 
         
         The \Bperp-dependence of the LPM emission mirrors that of Q1, as expected, with a diamagnetic shift of $\sigma = 1.00\ \pm$ \mueVTsq{0.04}, a negligible $\mu_{\perp}$ and the same DCP trend as Q1 (see Supplementary Figure 3). The similar magnetic field dependence of Q1 and LPM firmly supports the claim that the LPM emission is a replica of the same optical transition as Q1 with the additional emission of a phonon related to a local mode of the defect center\cite{Klein2021,Wigger2019}.
		
		\begin{figure}
			\includegraphics{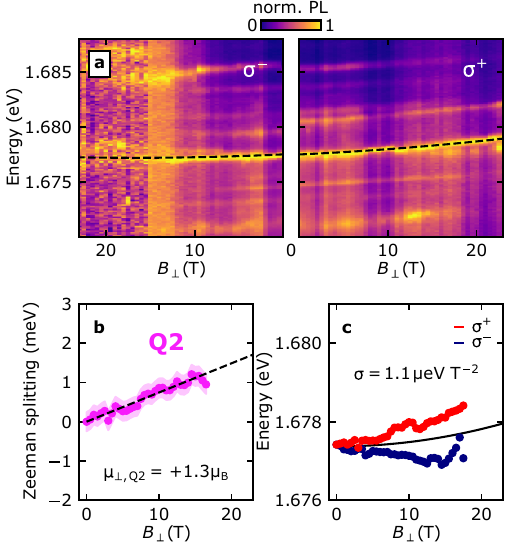}
% 			\internallinenumbers
			\caption{\label{Q2_Q3_Faraday}\textbf{Out-of-plane magnetic field \Bperpd dependent photoluminescence of defect luminescence Q2.} a) PL of Q2 versus \Bperpd for \circmind (left panel) and \circplusd polarized detection (right panel). b) The Zeeman splitting of Q2 shows a Zeeman splitting of $\mu_{\perp} = +1.3 \pm 0.1\ \mu_\text{B}$. c) Position of the ZPL of Q2 versus the applied magnetic field (\Bperp ), showing a diamagnetic shift of $\sigma = 1.1 \pm \mueVTsq{0.2}$.}
		\end{figure}
		
		We now turn to the magneto-spectroscopy of the emission line Q2. Figure \ref{Q2_Q3_Faraday}a shows polarization resolved spectra of the quantum emission Q2 as a function of magnetic field. 
	    We focus on the dominant peak at \eV{1.677}, and note that fainter peaks observed at this particular spot on the sample shift equally with \Bperpd (see Supplementary Figure 6 for magneto-spectroscopy of more locations). 
		Unlike Q1, we observe a sizeable valley Zeeman shift. This observation necessitates the lifting of valley degeneracy with increasing magnetic field and points towards optical transitions at the $K/K'$-points. The extracted valley Zeeman splitting is depicted in Fig. \ref{Q2_Q3_Faraday}b and yields a positive magnetic moment $\mu_{\perp}= +1.3 \pm 0.1\ \mu_\text{B}$, a sign that is opposite to the neutral exciton of the host \moly. 
		Furthermore, similar to Q1, we observe a diamagnetic shift with $\sigma = 1.1 \pm \mueVTsq{0.2}$ (see Fig. \ref{Q2_Q3_Faraday}c), indicating again that this emission originates from a bound state. 
		The magnetic field dependent DCP of Q2 shows a different behaviour as Q1 and can be fit with the usual equation using Boltzmann statistics of the involved spin-split defect states of cD1 (see Fig. \ref{Quantum_emission_overview}d and Supplementary Figure 5).
		The diamagnetic shift of Q* is negligible (see Supplementary Figure 7-9). In turn we interpret its emission to stem from a wave function of chemically functionalized defects.

		%The third category of quantum emission, Q3, is $\approx \meV{60}$ blue-shifted from Q1. Figure \ref{Q2_Q3_Faraday}d shows polarization resolved spectra for a representative emission at \eV{1.810} %showing what?. 
		%Unlike Q1 and Q2, we do not observe a diamagnetic shift at any of the investigated sites across the sample (see Figure \ref{Q2_Q3_Faraday}e and SI). Like Q2, we observe a valley Zeeman shift, however the sign of the g-factor $g_{\perp,Q3} = -1.02 \pm 0.10$ is negative.
		%As such, the observations on Q3 suggest a band-to-band or defect to defect transition where the magnetic moment of the involved bands is different from each other ($\mu_c \neq \mu_v$). 
		
		%Maybe this at the very beginning of the text about B-field measurements.
		%We measure the above characteristics for Q1, Q2, and Q3 on eight sites across the sample (see Supplementary information). The magnetic field induced shifts are consistent on all positions for the most abundant quantum emission Q1. Five of these show additional quantum emission of kind Q2. The emission Q3 can be found on three sites. The lack of signal for Q2 and Q3 can be explained with a lower oscillator strength of the corresponding optical transition, discussed below. 
		
		%In Figure \ref{Q2_Q3_Faraday}h we show the DCP as a function of \Bperpd for Q3. We observe a similar behaviour in magnetic field as for Q1 and Q2. \perc{50} of circular polarization is reached at \T{14.01}. The fitted parameters for Equation \ref{DCP_fit} are $E_{ex} = 1.33 \pm$ \meV{0.12} and $g = -1.65 \pm 0.15$.

	\subsection{\label{Voigt_geometry}In-plane magnetic field measurements on defects in \moly}
	
		\begin{figure}
			\includegraphics{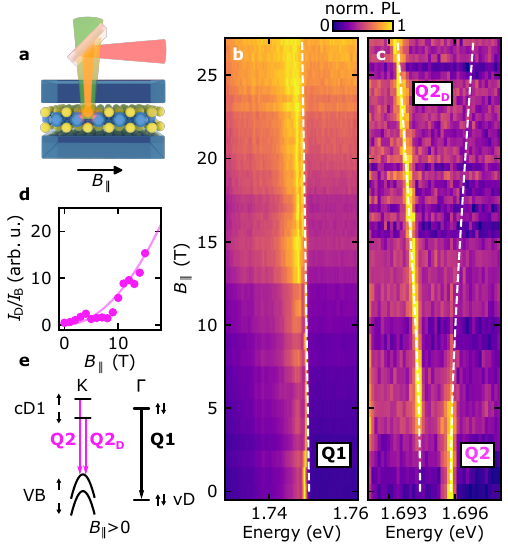}
% 			\internallinenumbers
			\caption{\label{Q1_Q2_Q3_Voigt}\textbf{In-plane magnetic field \Bpard measurements on the photoluminescence of defect luminescence Q1 and Q2.} a) Sketch of the in-plane magnetic field \Bpard configuration (Voigt Geometry). No polarization optics were used in the detection path. b) Emission of Q1 remains bright for all fields. c) The in-plane field reveals a second state Q2$_\text{D}$ energetically below Q2. The white dashed lines are guides to the eye for the expected in-plane Zeeman shift of Q1 and the dark-bright-splitting of Q2. The dark-bright splitting  $\Delta_\text{DB} = \meV{1.4}$ for Q2 is calculated with equation (\ref{dark-bright_splitting}). d) The quadratic dependence of the emission ratio between Q2$_\text{D}$ and Q2 indicates the brightening of a dark ground state. e) Sketch of the defect levels with the possible optical transitions for Q1 and Q2 at finite \Bpar.}
		\end{figure}
		
		To further investigate the details of the electronic states involved in the emission lines Q1 and Q2, we turn to magneto-spectroscopy in the Voigt configuration for which the magnetic field (\Bpar ) is applied parallel to the sample plane and perpendicular to the optical beam path (see Fig. \ref{Q1_Q2_Q3_Voigt}a and Supplementary Note 3 for data on Q*). In monolayer TMDs, strong spin-orbit coupling leads to out-of plane spin eigenstates particularly at the $K/K'$ points in the Brillouin zone. In principle, an in-plane magnetic field \Bpard induces a precession of the out-of-plane spins, leading to an increased mixing of the spin-eigenstates with increasing \Bpar. This mixing results in a magnetic-field dependent brightening of spin-forbidden transitions at $K/K'$ with the intensity of the dark transitions increasing relative to the bright transitions with \Bpar$^2$\cite{Molas2017, Robert2020, Kapuscinski2021}, similar to dark-bright mixing of interband transitions in semiconductor quantum dots with $C_{3v}$ symmetry\cite{Sallen2011}.
		%We collect the reflected light without polarization analyzer, since we do not expect the transitions appearing in PL to be valley-selective for an in-plane magnetic field. \cite{Sallen2012}  
		Figure \ref{Q1_Q2_Q3_Voigt}b shows a colormap of the PL versus \Bpard for Q1. We observe a monotonous redshift of the emission line, which is accompanied with spectral jitter with increasing magnetic field. Importantly, no brightening of a lower lying emission line is observed. The dashed white line in Fig. \ref{Q1_Q2_Q3_Voigt}b is a guide to the eye that shows a redshift of Q1 with an in-plane magnetic moment $\mu_{\parallel}= -1\ \mu_\text{B}$. 
		
		The in-plane magneto-PL of a typical Q2 quantum emitter is shown in Fig. \ref{Q1_Q2_Q3_Voigt}c. At \Bpard = \T{0}, one emission line is observed at \eV{1.696}. This line diminishes with increasing \Bpar, while a new peak, \meV{1.4} red-shifted from the original Q2 emission line, quickly appears above \T{3}. Figure \ref{Q1_Q2_Q3_Voigt}d shows the integrated PL ratio of the two lines together with a quadratic-in-$B$ fit, consistently describing the brightening of a dark transition. The low energy peak continues to red-shift with increasing \Bpard. This behavior can be described with the magnetic field induced splitting of a dark and bright emission branch\cite{Lu2020}
		\begin{eqnarray}
            \Delta\text{(} B_{\parallel}\text{)} =
            \Big{(} \Delta_\text{DB} \pm \sqrt{\Delta_\text{DB}^{2} + \text{(}\mu_{\parallel} \cdot B_{\parallel} \text{)}^{2}} \Big{)}
            \label{dark-bright_splitting},
        \end{eqnarray}
		where $\Delta_\text{DB}$ is the dark-bright splitting at \Bpard = \T{0} and $|\mu_{\parallel}|$ is the magnitude of the in-plane magnetic moment. The white dashed line in Fig. \ref{Q1_Q2_Q3_Voigt}c depicts equation (\ref{dark-bright_splitting}) with $\Delta_\text{DB} =$ \meV{1.4} and $|\mu_{\parallel}|= 1\ \mu_\text{B}$. The excellent agreement between data and fit together with the quadratic dependence of the relative intensities of dark and bright emission branch (see Fig. \ref{Q1_Q2_Q3_Voigt}d) shows the brightening of a dark transition.
		This observation necessitates the involvement of two spin states in Q2, while the observation of the valley Zeeman splitting shown above requires the breaking of valley degeneracy. As such, we conclude that Q2 must be a superposition of transitions involving both cD1 states and the valence band of the host \molyd (see Fig. \ref{Q1_Q2_Q3_Voigt}e).
		For neutral excitons in TMDs, a redshift with increasing \Bpard was explained with the average valley Zeeman shift of a bright and dark state ($|\mu_{\perp\text{,dark}}|-|\mu_{\perp\text{,bright}}|=2|\mu_{\parallel}|$)\cite{Molas2017,Lu2020,Robert2020}. Our observed shift is in very good agreement with the calculated difference of the bright and dark transition of Q2, which is simply given by the Zeeman splitting of the cD1 band ($\Delta\mu_\text{cD1} = 2.2 \mu_\text{B}$), extracted from Fig. \ref{Q1_LPM_Faraday}e. Finally, the combined magneto-spectroscopy in the Faraday and Voigt geometry unambiguously identifies Q2 as a spin conserving transition from the cD1 state to the respective \molyd valence band in the $K/K'$ valley.
		
		The fact that the emission line Q1 is energetically higher than Q2, while the diamagnetic shift, and therefore the binding energy of Q1 and Q2 are essentially the same, requires the defect band vD to be located below the valence band edge of \molyd, as sketched in Fig. \ref{Q1_Q2_Q3_Voigt}e\cite{Refaely-Abramson2018, Mitterreiter2021, Gupta2019}.    
		%The spectral weight of the PL will eventually come solely from the brightened line at high \Bpar, since it lies lower in energy. 
		%Figure \ref{Q1_Q2_Q3_Voigt}d,f depict the PL of the quantum emission Q3 in \Bpar. Again, a single emission line that is subject to spectral jittering between two states, appears at \Bpard = \T{0}. For a magnetic field \Bpard = \T{7} a second emission starts to brighten, roughly \meV{4} below. Interestingly, the spectral jitter of both peaks is synchronized, which substantiates the argument that the emission of both peaks originate from the same emission site (see Supplementary information). The white dashed line in Figure \ref{Q1_Q2_Q3_Voigt}d is a simulation using Eq. \ref{dark-bright_splitting} with $|$\gpar$|$ $ = 2$ and $\Delta_{db,0} =$ \meV{4}. Like Q2, we observe a quadratic increase of the integrated intensity with increasing \Bpar. 
		%The observed in-plane magnetic field behavior of Q2 and Q3 is consistent with the transitions originating at the K/K'-points of the Brillouin zone, where the strong spin-orbit coupling dictates an out-of-plane spin eigenstate. The observed in-plane Zeeman shift of all lines suggest a similar wave function symmetry of the electron, independent on the position in momentum space. 

	\subsection{\label{first-principle_calcluations}First-principles calculations on monolayer \molyd with embedded sulfur vacancies}

		\begin{figure*}
			\includegraphics{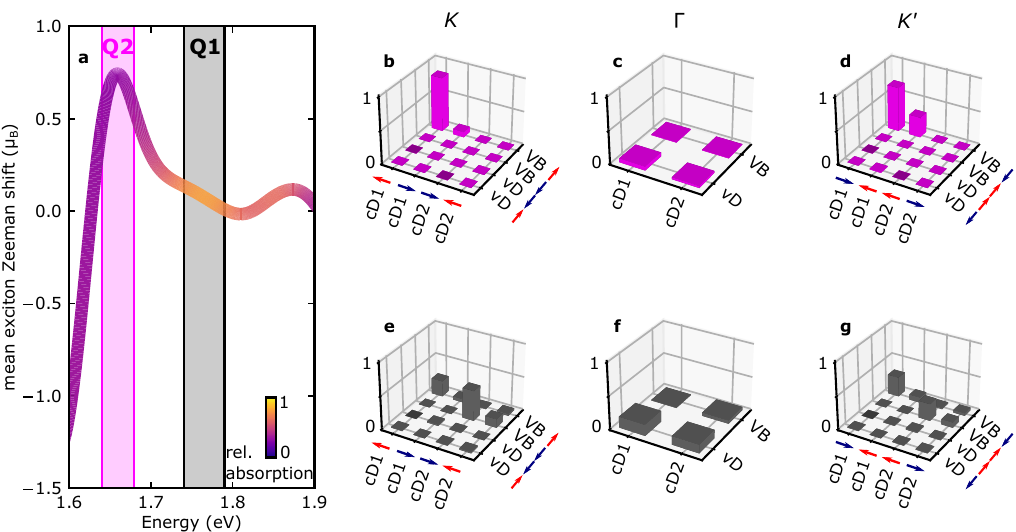}
% 			\internallinenumbers
			\caption{\label{Theory}\textbf{GW-BSE results for optical absorption, exciton Zeeman splitting, and transition contributions.} a) Mean exciton Zeeman splitting as a function of excitation energy. The colorcode on the line represents the absorption strength at each energy. b,c,d) Electron-hole transitions contributing to excitons in the energy range associated with Q2, namely in the magenta shaded area in (a). The electron bands (cD1 and cD2), as well as the hole bands (vD and VB) are illustrated in energetic order under the consideration of the lifted spin degeneracy at the $K/K'$-points. e,f,g) Electron-hole transitions contributing to excitons in the energy range associated with Q1, namely in the grey shaded area in (a). The transitions are shown for three selective k-points, $K$ (b,e), $\Gamma$ (c,f), and $K'$ (d,g).}

		\end{figure*}
		
		To deepen the insight into the nature of Q1 and Q2, we theoretically investigate exciton transitions in monolayer \molyd with embedded sulfur vacancies using ab-initio calculations.
		Details about the calculation can be found in the Supplementary Note 4 as well as the Methods section of \cite{Refaely-Abramson2018, Mitterreiter2021, Amit2022}.
		In brief, we use many-body perturbation theory within the GW-Bethe Salpeter approximation with explicit spin-orbit coupling and spinor wave functions and compute the many-body Zeeman splitting following \cite{Deilmann2020}.
		Figure \ref{Theory}a shows the mean calculated Zeeman splitting for an excitonic absorption as a function of excitation energy. Line colors represent the calculated absorption strength for \circplusd polarized light. The mean Zeeman splitting at each energy is calculated by averaging the Zeeman splitting of all discrete excitons composing the overall optical absorption in a narrow energy band. The transitions are broadened with a Gaussian and weighted by the oscillator strength of the respective excitons, which is calculated with the mentioned first-principles methods, accounting for the coupling between the associated electron and hole wave functions upon interaction with light (see Supplementary Figure 10). It has previously been shown\cite{Refaely-Abramson2018, Mitterreiter2021, Amit2022}, that the energetically lowest interband transitions are mainly composed of pristine \molyd valence band to defect band transitions. The strongly varying average exciton Zeeman splitting is testament of approximately conserved valley selectivity in this energy range, which sensitively depends on the distinct electron-hole transitions in a specific energy interval. We observe strong variations of the exciton Zeeman splitting in the energy range below $\sim \eV{1.70}$, originating from the hybridization of the BSE excitations, which mix electron-hole transitions from defect and non-defect bands. Specifically in the range of $\sim \eV{1.64}-\eV{1.68}$ we observe a significant increase of the Zeeman splitting, consistent with our observation for the Q2 emission.
		This is a result of allowed transitions to in-gap defect states of both spin components (see Fig. \ref{Quantum_emission_overview}c)
		In Fig. \ref{Theory}b,c and d we show the normalized contributions of electron-hole transitions composing the excitons in this energy range (magenta shaded area in Fig. \ref{Theory}a) at three representative points in the Brillouin zone ($K$, $\Gamma$, and $K'$). The height of each bar corresponds to the relative contribution of transitions from an occupied state (x-axis) to an unoccupied state (y-axis) upon excitation with \circplusd polarized light.
		At the $K/K'$ points, we find strong contributions for electron-hole transitions between the pristine-like \molyd valence band and both spin states of cD1. As a result, in the region of positive average Zeeman splitting, the contribution is highest for transitions from the upper \molyd valence band to the defect state cD1, as contributions at the $\Gamma$-point are comparatively reduced.
		For Q1, the energy range between $\sim \eV{1.74}-\eV{1.79}$ is selected. In this energy range, the absorption strength dominates, which is consistent with the dominating PL of Q1 as compared to the other defect emission bands. Furthermore, a slightly positive exciton Zeeman splitting is calculated, which again is consistent with our observations in Fig. \ref{Q1_LPM_Faraday}d. Figure \ref{Theory}e,f and g show the contributions of electron-hole transitions composing the exciton in the corresponding energy range. At the $K/K'$-points, only band-to-defect transitions are contributing, whereas at the $\Gamma$-point defect-to-defect transitions are dominating. Specifically, we find contributions from the lower valence band to defect band cD2 at the $K/K'$-points in absorption.
		Unlike Q1, the calculations suggest that the Q2 emission is only comprised of band-to-defect transitions at the $K/K'$-points. The hybridization of Q1 with the energetically lower Q2 emission can be deduced from the band-to-defect contributions at the $K/K'$-points.

\section{\label{Discussion}Discussion}

	%In conclusion we present magneto-spectroscopy of distinct quantum emitters based on sulfur vacancies in a host crystal monolayer \molyd.

% 	DISCUSSION: List the experimental observation and say that this is consistent with Theory.
	In order to characterize the defect luminescence, Q1 and Q2 of He-ion irradiated monolayer \molyd towards spin-defect properties, we combine our experimental observations with theoretical insight. He-ions create sulfur vacancies, which induce flat defect bands throughout the pristine bandstructure of monolayer \moly. As a result, a high joint DOS for defect-to-defect transitions is created. At the $K/K'$-points however, the defect vD lies below the valence band maximum of \moly, such that defect-to-band transitions are more likely.
	In high-field magneto-spectroscopy, we observe a diamagnetic shift of the defect luminescence Q1 and Q2, which is consistent with a bound particle of $\sim \nm{3.5}$. This localized character will result in optical transitions covering a significant momentum-space. Thus, the GW-Bethe-Salpeter equation is useful for gaining deeper understanding into the defect-induced transitions. Here we find that the nature of Q1 and Q2 can be viewed as mixed states of defect-to-defect and defect-to-band transitions. The level of this admixture determines the magnetic moment and valley selectivity of these hybridized transitions\cite{Refaely-Abramson2018, Amit2022}. 
	We find a dominant contribution of band-to-defect transitions at the $K/K'$-points for the energy interval of Q2, whereas Q1 additionally acquires sizeable defect-to-defect character from the $\Gamma$-point. These contributions at the $\Gamma$-point induce a breaking of the valley selectivity, reflected in the small valley Zeeman splitting of this transition. 
	For both Q1 and Q2, however, we find a magnetic-field dependent DCP, which necessitates spin conserving optical transitions. The distinct behavior of the DCP is explained by the defect-to-band transitions at the $K/K'$-points contributing to Q1 and Q2. 
	The in-plane magneto-spectroscopy reveals a spin-forbidden dark ground state for Q2, which unambiguously proofs the lifted spin degeneracy of the defect bands at the $K/K'$-points even at zero magnetic field. In contrast, the absence of a dark state for Q1 can be explained by the significant portion of transitions happening at the $\Gamma$-point, where the defect bands are spin degenerate due to Kramers' theorem.
% 	CONCLUSION:

	In conclusion, the combination of in-plane and out-of-plane magneto-spectroscopy identifies the Q1 emission as a defect-to-defect transition with admixture of Q2, which is dominated by transitions from in-gap defect states to the pristine valence band of \moly. This outcome suggests tailored modification of the defect luminescence through either charging or chemical modification (Q*, see Supplementary Note 2 and 3). We show that the defect states at $K/K'$ are split at zero magnetic field, a property that characterizes the sulfur vacancy in \molyd as a spin defect with desirable features for possible quantum technological applications.

\section{\label{Methods}Methods}

	\subsection{\label{Sample preparation}Sample preparation}
	 \molyd bulk crystals were purchased from HQGraphene, and the hBN crystals were provided by Takashi Taniguchi and Kenji Watanabe from NIMS, Japan. Monolayers of \molyd and few-layer hBN were obtained by mechanical cleavage of bulk crystals. The thin crystals were stacked with the viscoelastic transfer method onto a Si substrate with \nm{285} of thermal \sildiox. After the assembly of the desired heterostructure, namely the \molyd encapsulated in hBN, the helium ion microscope (HIM) Orion NanoFab from Zeiss was used to precisely irradiate the sample with He-ions at \kV{30} in an array pattern with a pitch of \mum{2}. The dose was chosen to get a high yield of single sharp emitter lines for Q1. For the photoluminescence measurements in Figure \ref{Quantum_emission_overview}b an excitation wavelength of \nm{639} and a power of \nW{550} at a bath temperature of \K{1.7} was used. The photoluminesence signal was guided on a nitrogen cooled CCD via a 300 grooves/mm dispersive grating. A long pass filter was used to extinguish the directly reflected excitation laser emission.
	
	\subsection{\label{Magneto-optical measurements}Magneto-spectroscopy}
	The magneto-photoluminescence measurements were performed in a cryostat cooled to \K{4.2} surrounded by a resistive magnet. For excitation, a laser diode with an emission wavelength of \nm{515} was used. For the measurements in Faraday configuration (B-field perpendicular to the sample plane and parallel to the optical beam path), the linearly polarized excitation was focused on the sample with an objective with NA = 0.81. The reflected light was collected with the same objective and guided through a $\lambda/4$-plate followed by a linear polarizer to select the \circplusd (\circmin ) polarized light in the detection. A long pass filter was used to extinguish the directly reflected excitation laser emission. The collected light was analyzed in a spectrometer equipped with a liquid nitrogen cooled CCD and a 600 grooves/mm dispersive grating. For measurements in Voigt configuration (B-field parallel to the sample plane and perpendicular to the optical beam path) the sample was mounted vertically and the detection was unpolarized. A mirror tilted by 45° was used to guide incident light perpendicular to the sample plane. A large working distance objective with NA = 0.35 was used to focus the excitation laser via the tilted mirror onto the sample and collect the reflected light.
	
	\subsection{\label{Ab-initio Calculations}Ab-initio calculations}
    State-of-the-art ab initio ground-state and excited-state calculations were carried out in a $5\times5\times1$ supercell of monolayer \molyd composed of 74 atoms and a single sulphur vacancy. Ground state DFT calculations were performed using the Quantum Espresso package\cite{Giannozzi2009,Giannozzi2017} for assessing the atomic structure, spinor wave functions and single-particle magnetization, with an energy cutoff of 75 Ry. These properties were used as a starting point for a GW\cite{Hybertsen1986} calculation of the quasi-particle energies, including spin-orbit coupling within the BerkeleyGW\cite{Deslippe2012} software, with summation over 3998 bands on a $3\times3\times1$ k-grid and an energy cutoff of 25 Ry for the dielectric matrix. Electron-hole coupling and exciton energies were calculated using BerkeleyGW by solving the Bethe-Salpeter equation (BSE)\cite{Rohlfing1998,Rohlfing2000} by interpolating the GW results to a k-grid of $6\times6\times1$ with a dielectric matrix which was calculated with a 5 Ry cutoff and summation over 1798 bands. Quasi-particle magnetization corrections and many-body excitonic Zeeman splitting were evaluated from the GW-BSE results following recently derived methods\cite{Wozniak2020,Deilmann2020}. More details are presented in the Supplementary Note 4.

\section{Data Availability}
Data included in this manuscript will be made available upon reasonable request to the authors.

\bibliographystyle{npj}
\bibliography{main}% Produces the bibliography via BibTeX.

% \printbibliography

\begin{acknowledgments}
The work was supported by Deutsche Forschungsgemeinschaft (DFG). We gratefully acknowledge financial support of the German Excellence Initiative by MCQST (EXS-2111) and e-conversion (EXS-2089). This work has been partially supported by the EC Graphene Flagship project and by ANR projects ANR-17-CE24-0030 and ANR-19-CE09-0026. This work was supported by LNCMI-CNRS, members of the European Magnetic Field Laboratory (EMFL). J.K. acknowledges support by the Alexander von Humboldt foundation. K.W. and T.T. acknowledge support from the JSPS KAKENHI (Grant Numbers 19H05790, 20H00354 and 21H05233). T.A, G.C., D.H., and S.R-A. acknowledge support from the David Lopatie Fellows Program and the ERC Starting grant 101041159. S.R. acknowledges support from the Independent Research Fund Denmark.
\end{acknowledgments}

\section*{\label{author_contributions}Author Contributions}
    A.H., J.K., J.J.F., A.W.H., and A.S. conceived and designed the experiments. T.A, G.C., D.H., and S.R-A. performed the DFT-GW and BSE calculations. S.R., J.K. prepared the sample. K.W. and T.T. provided high-quality hBN bulk crystals. A.H., A.S., T.P., A.D., C.F., J.K. and K.B. performed the optical measurements. A.H. analyzed the data. C.K., M.P. and C.F. contributed interpreting the data. A.H. and A.S. wrote the manuscript with input from all coauthors.

\section*{\label{competing_interests}Competing interests}
    The authors declare no competing interests

%----------------------------SI-------------------------------
\newpage
\onecolumngrid
\begin{flushleft}
Supplementary Information
\end{flushleft}

\section*{\label{sample_characterization}Supplementary Note 1: Sample characterization}
    
    \begin{figure*}[ht!]
		\includegraphics{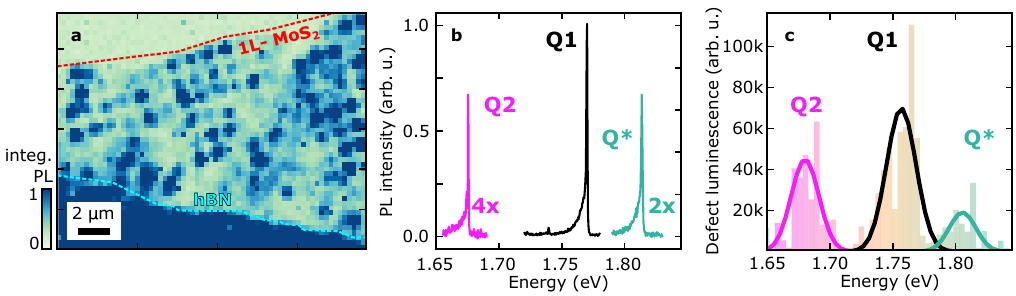}
% 		\internallinenumbers
		\caption{\label{PL_distribution}\textbf{Spatial and energetic distribution of the defect luminescence.} a) Spatial map of the defect luminescence integrated from 1.72 to \eV{1.78}. The monolayer \molyd is located below the red dashed line and is encapsulated in hBN from the turquoise dashed line upwards. Broad backgrounds in the photoluminescence spectra are substracted. Scale bar, \mum{2}. b) Typical low-temperature photoluminescence (PL) spectra of the quantum emission Q1, Q2, and Q*. c) Spectral position of the defect luminescence weighted by intensity. Three distinct emission bands are visible. The most prominent is located at $\sim \eV{1.75}$ (Q1). Red-shifted to that, at $\sim \eV{1.69}$ is the emission band of Q2. The blue-shifted emission band Q* is located at $\sim \eV{1.81}$.}
	\end{figure*}
	
	The heterostructure consists of a monolayer \molyd encapsulated in multi-layer hexagonal boron nirtide (hBN). The three layers of 2D materials are positioned upon each other by dry viscoelastic stamping methods. Supplementary Figure \ref{PL_distribution}a shows a spatial false color plot of the photoluminescence (PL) integrated from 1.72 to \eV{1.78}. The red dashed line illustrates the upper boundary of the \molyd flake, whereas the dashed blue line represents the lower boundary of the top hBN flake. The heterostructure, located in between these boundaries, was subsequently irradiated with He-ions. The irradiated pattern is an array of circular patches with a pitch of \mum{2}. Optically active defects are generated at the irradiated sites, which appear as sharp emission lines in PL measurements (see Fig S\ref{PL_distribution}b). The emission lines are distributed in three distinct clusters Q1, Q2 and Q* (see Supplementary Figure \ref{PL_distribution}c). We attribute emission lines from the later cluster, namely Q*, to adsorbate related defect luminescence\cite{Klein2021} and therefore exclude its evaluation from the main manuscript.

    \begin{figure*}[ht!]
		\includegraphics{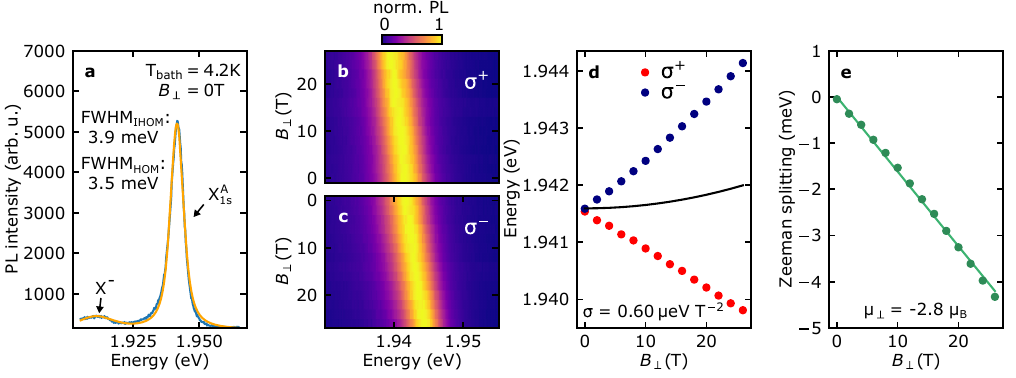}
% 		\internallinenumbers
		\caption{\label{neutral_exciton}\textbf{Valley-Zeeman splitting of the neutral exciton in monolayer \moly.} a) PL spectrum of the neutral exciton  with a Voigt fit for exciton $X_{1s}^{A}$ and trion $X^{-}$ with an inhomogeneous (\meV{3.9}) and a homogeneous part (\meV{3.5}) of the FWHM at \Bperpd = \T{0} and a bath temperature $T_\text{Bath}$ = \K{4.2}. b) - c) Normalized photoluminescence (PL) of the neutral exciton in \circplusd and \circmind detection, respectively. d) Diamagnetic shift of the neutral exciton with $\sigma = 0.60\ \pm$ \mueVTsq{0.1}. e) Valley Zeeman splitting of the neutral exciton with $\mu_{\perp} = -2.8 \pm 0.1\ \mu_\text{B}$.}
	\end{figure*}
	
	Supplementary Figure \ref{neutral_exciton}a shows a typical spectrum of the neutral exciton $X_{1s}^{A}$ of \moly. A trion peak appears $\sim \meV{30}$ red-shifted to the exciton. We find an inhomogeneous and homogeneous broadening of the lineshape with \meV{3.9} and \meV{3.5}, respectively. The \Bperpd dependent spectra are illustrated as false color plots in Supplementary Figure \ref{neutral_exciton}b (c) for \circplusd (\circmin ) polarized emission. By fitting the spectral positions we find a diamagnetic shift of $0.6 \pm \mueVTsq{0.1}$ (Supplementary Figure \ref{neutral_exciton}d) and a valley Zeeman splitting of $\mu_{\perp} = -2.8 \pm 0.1\ \mu_{B}$ (Supplementary Figure \ref{neutral_exciton}e).

\section*{\label{additional_out_of_plane_data}Supplementary Note 2: Additional out-of-plane magnetic field data}

	\begin{figure*}[ht!]
		\includegraphics{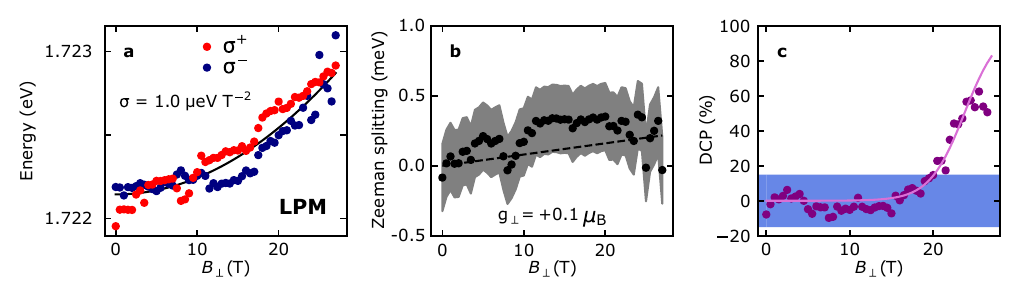}
% 		\internallinenumbers
		\caption{\label{LPM}\textbf{Magneto-spectroscopy on LPM feature.} a) Average spectral position of LPM versus \Bperpd for \circplusd (red) and \circmind (blue) polarized detected light. A diamagnetic shift of $\sigma = 1.0 \pm$ \mueVTsq{0.1} can be extracted. b) Valley Zeeman-splitting of LPM feature (\gperp $= +0.1 \pm 0.1\ \mu_\text{B}$). c) Degree of circular polarization (DCP) of LPM fitted with equation (1) from the main manuscript.}
	\end{figure*}
	
	About \meV{30} red-shifted to the quantum emission Q1, we find an additional peak, which we attribute to a local phonon mode (LPM) of the defect center. The magneto-spectroscopy results of the LPM mirror the one of Q1 for the diamagnetic shift (Supplementary Figure \ref{LPM}a), the Zeeman splitting (Supplementary Figure \ref{LPM}b), and the degree of circular polarization (Supplementary Figure \ref{LPM}c).

	\begin{figure*}[hb!]
		\includegraphics{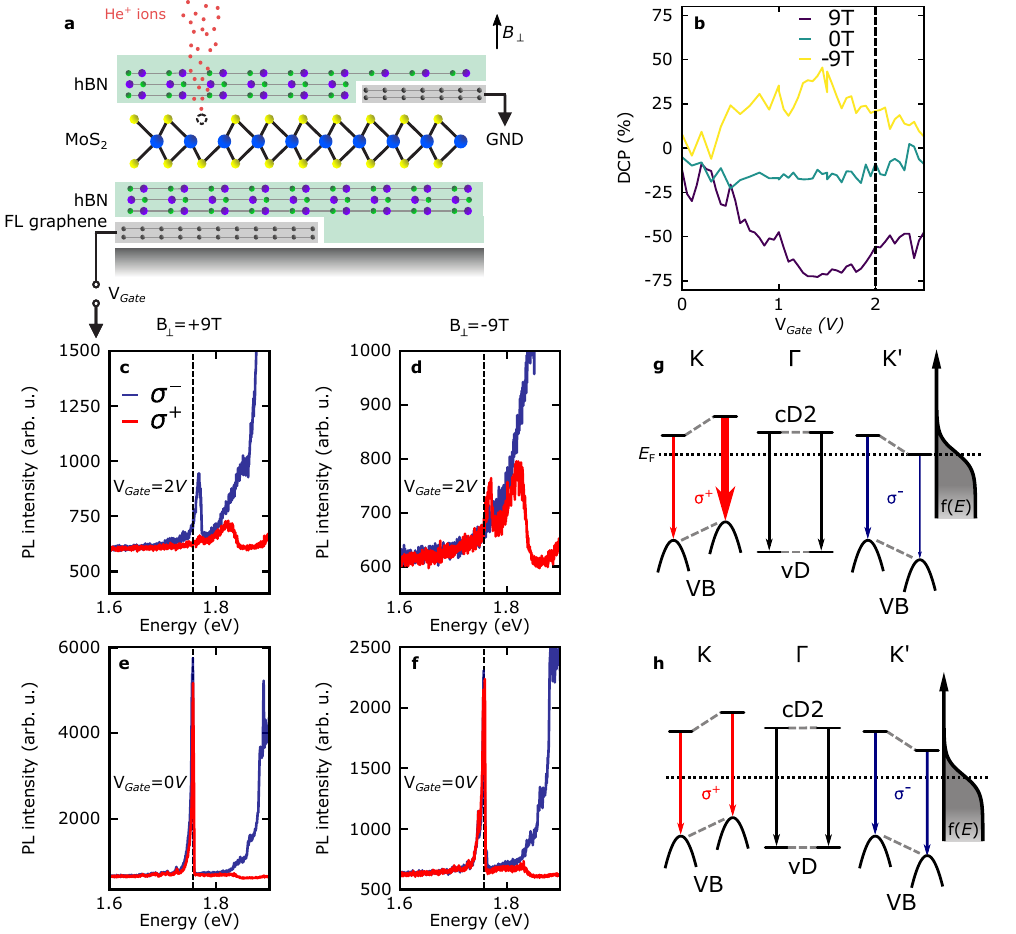}
% 		\internallinenumbers
		\caption{\label{gate_dep_DCP}(Figure caption on the next page.)}
	\end{figure*}
	\begin{figure*}[ht!]
	   % \internallinenumbers
	   % \captionstyle{\centerlastline}
	    \contcaption{\textbf{Degree of circular polarization (DCP) of a Q1-emitter by tuning the Fermi energy $E_\text{F}$ and an out-of-plane magnetic field \Bperp.} a) Sketch of the hBN/\moly/hBN heterostructure. Few layer (FL) graphene provides a gate voltage $V_\text{Gate}$ and contact to the electric ground (GND). b) DCP of Q1 measured as a function of $V_\text{Gate}$ for \Bperpd $=$  \T{0}, $\pm$\T{9}. c,d) PL spectra of Q1 detected in \circplusd (red) and \circmin-polarized (blue) configuration at $V_\text{Gate} =$ \V{2} and \Bperpd $=$  $\pm$\T{9}. e,f) Similar spectra at $V_\text{Gate} =$ \V{0} and \Bperpd $= \pm$\T{9}. The dashed line highlights the energy of the zero-phonon line at $V_\text{Gate} =$ \V{0}, indicating a clear blue shift of the emitter at finite $V_\text{Gate}$. Experimental parameters: $E_\text{Laser} =$ \eV{1.94}, $P_\text{Laser} =$ \muw{1}, spot size = \mum{1}, $T_\text{Bath} =$ \K{1.7}. g,h) Sketches of the bandstructure with possible transitions at the high-symmetry points $K$, $\Gamma$ and $K'$ for two gate voltages at zero and positive magnetic field. The sketched energy levels summarize the finding of b-f) as follows: The DCP is close to zero for zero gate voltage (and negative voltages) within the given experimental uncertainty [compare b and e,f]. This finding is consistent with the following interpretation: if the Fermi-energy is sufficiently below the unoccupied defect-states, the Zeeman-split states (cD1 and cD2) are equally occupied after an optical excitation and therefore the DCP is negligible [compare b]. At slightly positive gate voltages, the DCP switches sign for negative and positive magnetic fields [compare b and c,d]. This observation suggests that the Zeeman-energy is the relevant energy scale, and the DCP depends on the occupation of the defect bands. For large positive gate voltages, the DCP reduces again [compare b], which is consistent with the spin-split states becoming equally occupied. Importantly, for $V_\text{Gate} >$ \V{0}, the Q1 emission shifts to higher energies [compare dashed lines in c,d wrt. e,f]. This blue-shift indicates that in the investigated gate voltage range, the emitters are charge-neutral, while charged free exciton states typically red-shift (including trion and attractive polaron).\cite{Efimkin2018,Fey2020,Hotger2021} Trion emission from the entire detection spot in the vicinity of Q1 is observed as a background emission at $\approx 1.8 eV$ in c and d.}% Continued caption
	\end{figure*}
	
	\begin{figure*}[ht!]
		\includegraphics{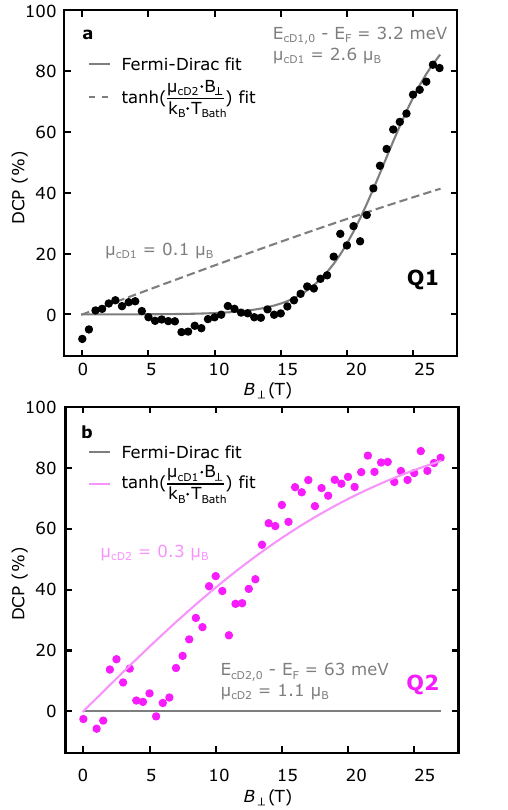}
% 		\internallinenumbers
		\caption{\label{fitting}\textbf{Fits of the DCP versus \Bperpd of Q1 and Q2.} a) The Fermi-Dirac fit (equation (1), mentioned in the manuscript) fits the DCP data of Q1 well. The fit parameters yield $E_\text{cD1,0}-E_\text{F} = \meV{3.2}$ and $\mu_\text{cD1} = 2.6\ \mu_\text{B}$. However, the $\tanh\Big{(}\dfrac{\mu_\text{cD2} \cdot \text{\Bperp}}{k_\text{B} T_\text{Bath}}\Big{)}$ fit is clearly not suitable to describe the data. b) For Q2 we assumed $E_\text{F}$ to be \meV{63} below $E_\text{cD1,0}$. The Fermi-Dirac fit cannot map the data points at all. Whereas the $\tanh\Big{(}\dfrac{\mu_\text{cD1} \cdot \text{\Bperp}}{k_\text{B} T_\text{Bath}}\Big{)}$ describes the DCP trend with a fitted value of $\mu_\text{cD1} = 0.27 \pm\ 0.01\ \mu_\text{B}$. Note, that this value is very sensitive to the bath temperature $T_\text{Bath}$, which might explain the discrepancy to the ab-initio calculations ($\mu_\text{cD1} = 1.07\ \mu_\text{B}$).}
	\end{figure*}
	
	\begin{figure*}[ht!]
		\includegraphics{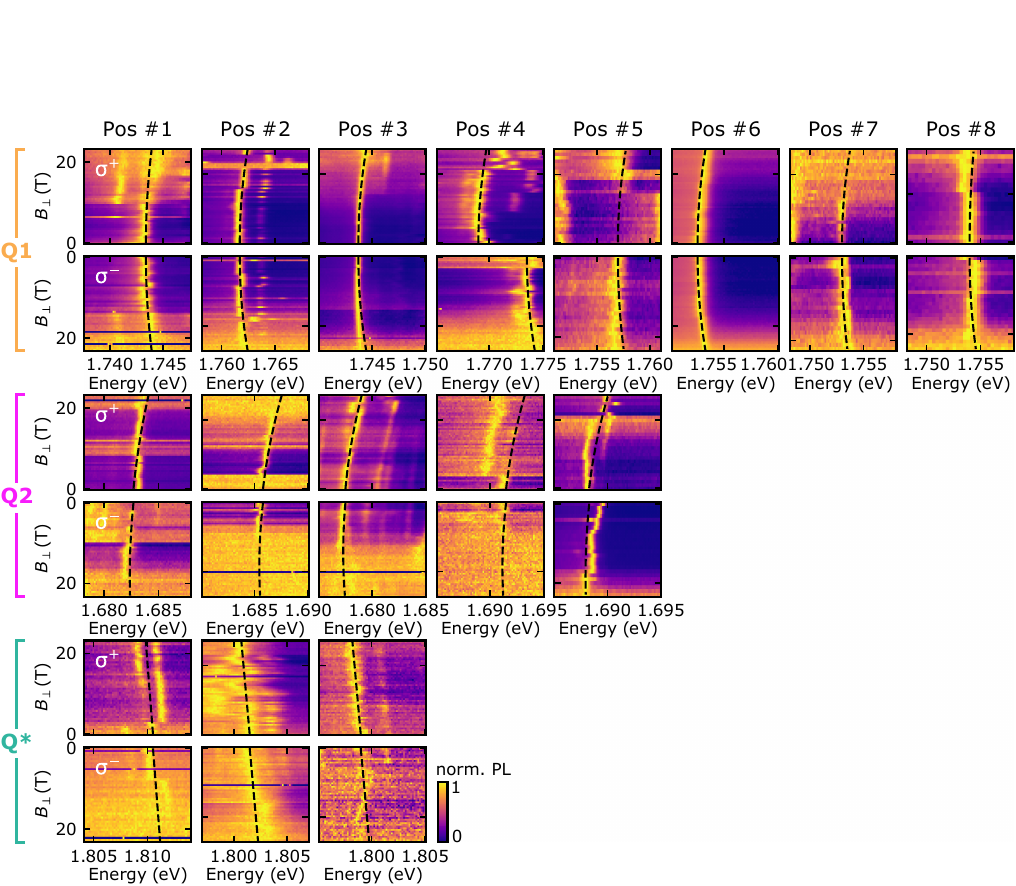}
% 		\internallinenumbers
		\caption{\label{statistics}\textbf{Overview of all magneto-PL measurements in Faraday geometry.} The black dashed lines are guides to the eye considering the fitted parameters found for Q1, Q2 and Q*, respectively. For all eight position an emission according to Q1 was found. Additionally at position $\#1 - \#5$, an emission conforming to Q2 was found. The quantum emission Q* was found at position $\#1 - \#3$. All emitter agree well with the assigned behaviour in magnetic field. Some emitters jitter between different states over time. Nevertheless, their relative shift in magnetic field remains unchanged.}
	\end{figure*}
	
	In total we measured at eight distinct positions on the sample (see Supplementary Figure \ref{statistics}). On three positions we were able to resolve all three emission lines Q1, Q2 and Q*. On five we were able to get signal from emission line Q1 and Q2 and on the remaining three we only saw Q1. The black dashed lines in the false color plots are guides to the eye for each quantum emission, which resemble Zeeman and diamagnetic shifts evaluated in the main manuscript for Q1, Q2 and Q*. The emitter at position $\#6$ is shown in the main manuscript for the evaluation on Q1. For Q2, we show position $\#3$, in the main manuscript. 
	
	\begin{figure*}[ht!]
		\includegraphics{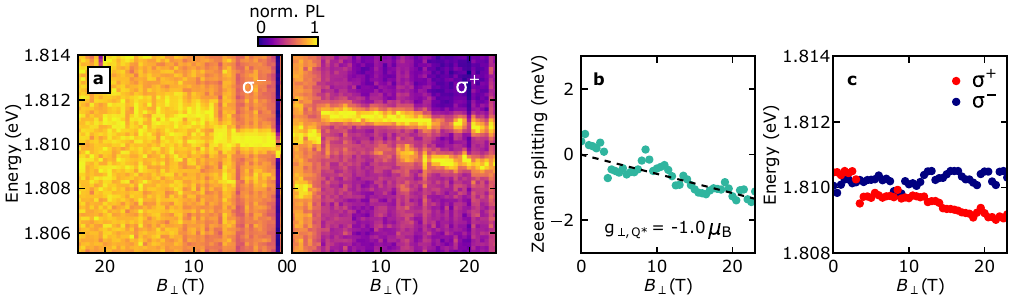}
% 		\internallinenumbers
		\caption{\label{Q_star}\textbf{Out-of-plane magneto-spectroscopy on Q* feature.} a) False color plot of the spectrum of Q* versus \Bperp. The ZPL of the defect luminescence is subject to jitter. b) Zeeman-splitting of Q* feature showing \gperp$_{\text{,Q*}} = -1.0 \pm 0.1\ \mu_\text{B}$. c) Average spectral position of LPM versus \Bperpd for \circplusd (red) and \circmind (blue) polarized detected light. We find no diamagnetic shift opposing to Q1 and Q2.}
	\end{figure*}
	
	Figure S\ref{Q_star}a) shows the out-of-plane magneto-spectroscopy of the Q* emission (Position $\#1$) in a false color plot. The ZPL of the Q* emission is subject to jitter. The evaluation of the Zeeman splitting is displayed in Supplementary Figure \ref{Q_star}b, which reveals a splitting of \gperp$_{\text{,Q*}} = -1.0 \pm 0.1\ \mu_\text{B}$. Within the measurement accuracy, we were not able to resolve a diamagnetic shift (see Supplementary Figure \ref{Q_star}c). It is reported that the encapsulation of the monolayer \molyd fairly reduces the luminescence of Q*. \cite{Klein2021}

\clearpage
\section*{\label{additional_in_plane_data}Supplementary Note 3: Additional in-plane magnetic field data}

	\begin{figure*}[ht!]
		\includegraphics{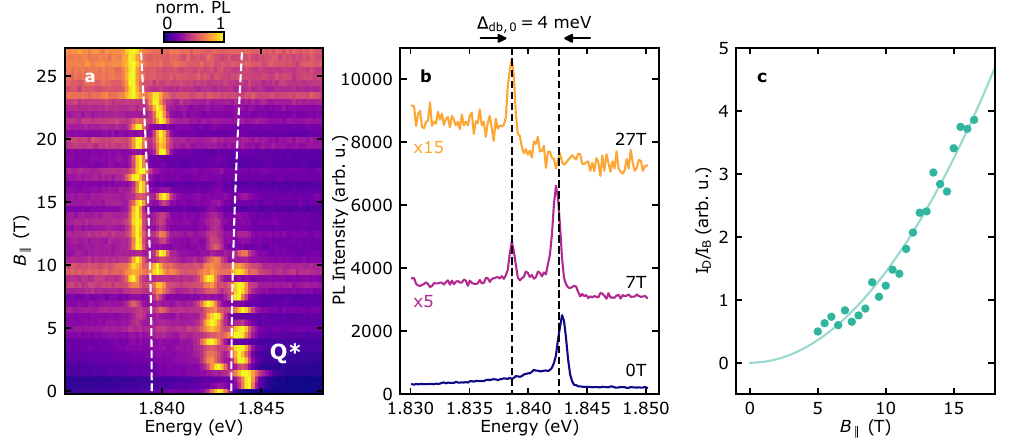}
% 		\internallinenumbers
		\caption{\label{Q_star_in_plane}\textbf{In-plane magneto-spectroscopy on Q* feature.} a) False color plot of the PL of Q* versus \Bpar. At \Bpard $\approx \T{5}$ a new peak red-shifted to Q* emerges, while the original peak faints out. The white dashed line is calculated with the dark-bright splitting equation mentioned in the main manuscript and uses $|\mu_{\parallel}| = 1\ \mu_\text{B}$ and $\Delta_\text{DB} = \meV{4}$. The defect luminescence is subject to jitter, which is further addressed in the Supplementary Figure \ref{time_trace_Q*}. b) Three distinct spectra at \Bpard $ = $ \T{0}, \T{7}, \T{27} showing the original, both, and the brightened defect luminescence, respectively. c) The ratio of the dark and bright emission intensity, reveals a quadratic dependence.}
	\end{figure*}
	
	The in-plane magnetic field \Bpard data on Q* reveals similar to Q2 a brightening of a spin-forbidden dark ground state (see Supplementary Figure \ref{Q_star_in_plane}). 
	
	\begin{figure*}[ht!]
		\includegraphics{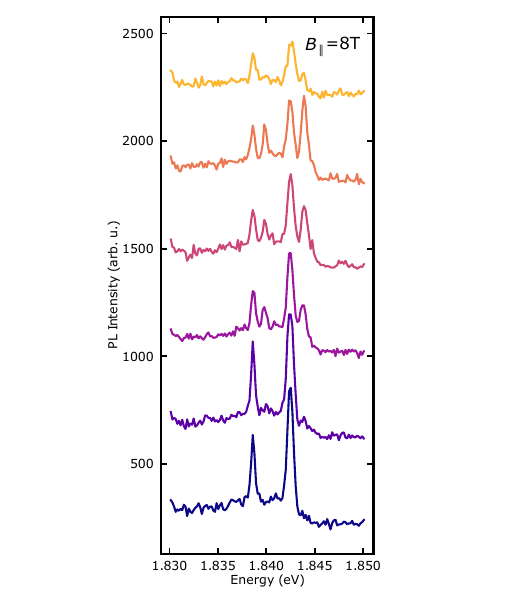}
% 		\internallinenumbers
		\caption{\label{time_trace_Q*}\textbf{Time trace of a dark and bright quantum emission Q* at constant in-plane magnetic field.} Waterfall plot showing six spectra taken one after the other at the same position at a fixed in-plane magnetic field \Bpard = \T{8}. The jitter of the blue-shifted peak, namely bright quantum emission, resembles the jitter of the dark quantum emission (red-shifted peak). This is another signature for the dark-bright splitting. The dielectric environment, which can vary drastically in real space, is mostly accountable for the jitter seen in the PL. Both emission lines must origin locally from the same position (i.e. seeing the same dielectric environment).}
	\end{figure*}
	
	The spectral jitter observable for the Q* emission, can be explained for instance by a dangling adsorbate on a sulfur vacancy. The new branch appearing at finite in-plane magnetic field \Bpard mirrors the spectral jitter behaviour, when keeping \Bpard constant (see Supplementary Figure \ref{time_trace_Q*}).

\clearpage
\section*{\label{theoretical_details}Supplementary Note 4: DFT, GW and BSE calculations}

    \begin{figure*}[h!]
		\includegraphics{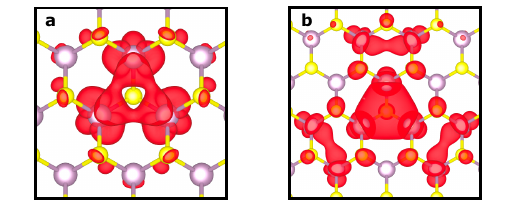}
% 		\internallinenumbers
		\caption{\label{defect_wavefunctions}\textbf{Isosurfaces of the defect wavefunctions.} Isosurfaces of the defect electron wavefunction corresponding to defect band a) cD1 and b) vD. The wavefunctions of cD1 and cD2 are similar and are primarily composed of transition metal d-orbitals.}
	\end{figure*}

    The spin magnetic moment of electronic band $n$ at the reciprocal-space point $\mathbf{k}$ was calculated using:

    \begin{equation} \label{spin_eq}
        m^{spin}_{n\mathbf k} = -\frac{e g_e}{2m_e}\langle \psi_{n\mathbf{k}} | \sigma_{z} | \psi_{n\mathbf{k}} \rangle,
    \end{equation}
    
    where $e$ is the elementary charge, $g_e$ is the free electron g-factor, namely Zeeman splitting, $m_e$ is the electron’s mass, $\psi_{n\mathbf k}$ is the wavefunction of band n at k-point $\mathbf{k}$ and $\sigma_z$ is the Pauli matrix in the $\hat{z}$ direction (perpendicular to the monolayer plane).
    
     The orbital magnetic moment was calculated using:
    \begin{equation} \label{orb_eq}
    m^{orb}_{n\mathbf k} = -\frac{i\mu_B}{m_e}\sum_{n'\ne n}{\Big{(}\frac{\left<\psi_{n\mathbf k}\left|\hat p_x\right|\psi_{n'\mathbf k}\right>\left<\psi_{n'\mathbf k}\left|\hat p_y\right|\psi_{n\mathbf k}\right>}{E_{n'\mathbf k}-E_{n\mathbf k}}-\frac{\left<\psi_{n\mathbf k}\left|\hat{p_y}\right|\psi_{n'\mathbf k}\right>\left<\psi_{n'\mathbf k}\left|{\hat p_x}\right|{\psi_{n\mathbf k}}\right>}{E_{n'\mathbf k}-E_{n\mathbf k}}\Big{)}}
    \end{equation}
    
    Where $\mu_B$ is Bohr’s magneton, $E_{n\mathbf{k}}$ is the energy of band $n$ at k-point $\mathbf{k}$ and $\hat{p_i}$ is the momentum operator at direction $\hat{i}$.
    % Because of degeneracies at the $\Gamma$ point which strongly affect this equation, the point presented in the main text is a point close to $\Gamma$ but not exactly $\mathbf{k} = (0, 0, 0)$.
    
    The spin and orbital parts of the magnetic moment are then added to calculate the total magnetic moment:
    \begin{equation} \label{total_eq}
    m^{tot}_{n\mathbf k}=m^{spin}_{n\mathbf k}+m^{orb}_{n\mathbf k}.
    \end{equation}
    
    We extend the single-particle results to a many-body excitonic picture, using the following equation to calculate the g-factor of each exciton based on the GW-BSE results for the excitonic compositions:
    \begin{equation}\label{BSE_eq}
        g^S = \frac{2}{\mu_B}\sum_{v c\mathbf k}{\left|A^{S}_{vc\mathbf{k}}\right|^2\big{(}m_{c\mathbf {k}}-m_{v\mathbf k}\big{)}},
    \end{equation}
    where $A^{S}_{vc\mathbf{k}}$ are the GW-BSE exciton coefficients, which weigh the effective-mass transitions.

% \bibliographystyle{npj}
% \bibliography{library}% Produces the bibliography via BibTeX.
%----------------------------SI-------------------------------

\end{document}